\ifx\pdfoutput\undefined
\documentclass[aps,prd,a4paper,twoside,twocolumn,english,superscriptaddress,
	showkeys,nofootinbib,showpacs,
	amssymb,amsmath,amsfonts]{revtex4}
\else
\documentclass[aps,prd,a4paper,twoside,twocolumn,english,superscriptaddress,
	showkeys,nofootinbib,showpacs,
	amssymb,amsmath,amsfonts]{revtex4}

\fi

\usepackage{babel}
\usepackage{amsmath}
\usepackage{amsfonts}
\usepackage{amssymb}
\usepackage{dsfont}

\begin{document}

\setcounter{page}{1}

\title{Fermions in Ashtekar-Barbero Connections Formalism for Arbitrary Values of the Immirzi Parameter}

\author{\firstname{Simone} \surname{Mercuri}}
\email{mercuri@icra.it}
\affiliation{Dipartimento di Fisica, Universit\`a di Roma ``La Sapienza'', P.le Aldo Moro 5, I-00185, Rome, Italy.}
\affiliation{ICRA --- International Center for Relativistic Astrophysics.}

\date{\today}

\begin{abstract}
The Ashtekar-Barbero-Immirzi formulation of General Relativity is extended to include spinor matter fields. Our formulation applies to generic values of the Immirzi parameter and reduces to the Ashtekar-Romano-Tate approach when the Immirzi parameter is taken equal to the imaginary unit. The dynamics of the gravity-fermions coupled system is described by the Holst plus Dirac action with a non-minimal coupling term. The non-minimal interaction together with the Holst modification to the Hilbert-Palatini action reconstruct the Nieh-Yan invariant, so that the effective action coming out is the one of Einstein-Cartan theory with a typical Fermi-like interaction term: in spite of the presence of spinor matter fields, the Immirzi parameter plays no role in the classical effective dynamics and results to be only a multiplicative factor in front of a total divergence. 

We reduce the total action of the theory to the sum of dynamically independent Ashtekar-Romano-Tate actions for self and anti-self dual connections, with different weights depending on the Immirzi parameter. This allows to calculate the constraints of the complete theory in a simple way, it is only necessary to realize that the Barbero-Immirzi connection is a weighted sum of the self and anti-self dual Ashtekar connections.
Finally the obtained constraints for the separated action result to be polynomial in terms of the self and anti-self dual connections, this could have implications in the inclusion of spinor matter in the framework of non-perturbative quantum gravity. 
\end{abstract}



\maketitle

\section{General Remarks}\label{par1}

In the last years with the introduction by Ashtekar of a new formalism for General Relativity \cite{ash86,ash87}, many steps forward have been made in finding a consistent quantum theory of gravity. The main advantage represented by the Ashtekar formalism in the program of a background independent quantization of the gravitational field is the reduction of the phase space of General Relativity (GR) to that of a Yang-Mills gauge theory, with the introduction of self-dual $SL(2,C)$ connection, satisfying appropriate reality conditions. Furthermore, introducing Ashtekar connections, the constraints of GR reduce to a polynomial form, opening the way to the canonical quantization procedure, which has led to the formalization of a background independent non-perturbative quantum theory of gravity, known as \emph{Loop Quantum Gravity} (LQG) \cite{rov04,ashlew04} (for a mathematically rigorous approach to LQG see \cite{thi01}, in order to deepen into the role of Wilson's loops in Quantum Field Theory and LQG see \cite{gam94}, arguments clarifying the meaning of background independence and of relational space-time are contained in \cite{smo05}, while a comparison between different approaches to the quantum gravity problem can be found in \cite{smo03}). 
\newline The difficulties connected with the implementation on the quantum level of the reality conditions led Barbero to introduce real $SU(2)$ connection \cite{bar95}, instead of the complex Ashtekar's one. In Barbero's formalism we have not any need of the reality conditions, but the Hamiltonian scalar constraint is more complicate than the one coming out using complex connections. The link existing between Ashtekar and Barbero connections was clarified by Immirzi \cite{imm97-1,imm97-2}, who observed that there exists a canonical transformation which allows to introduce a finite complex number $\beta\neq 0$, called the \emph{Immirzi parameter}, in the definition of the connection; this represents a generalization of the Ashtekar's formalism and reduces to the original one when the Immirzi parameter is taken equal to the positive or negative determination of the imaginary unit, $i$, corresponding to the self or anti-self dual Ashtekar variables; on the other hand real values of the Immirzi parameter yield the Barbero connections, originally defined taking $\beta=\pm1$. 

The Immirzi parameter is a free parameter of the theory and, being introduced, as said above, via a canonical transformation, does not affect the classical dynamics; but, as shown in \cite{rovthi98}, the canonical transformation cannot be implemented unitarily in the quantum theory, yielding striking effects, for example, in the spectra of the area and volume operators computed in LQG: they come out to be proportional respectively to $\beta\ell_{pl}^2$ and $\beta^{3/2}\ell_{pl}^3$.
\newline In this respect the role of the Immirzi parameter becomes clear when we treat the Ashtekar-Barbero-Immirzi formalism in the covariant Holst approach \cite{hol96}, it, in fact, results to be a multiplicative factor in front of a modification of the Hilbert-Palatini action; this modification does not affect the dynamics if space-time is torsion-less. Therefore the role of the Immirzi parameter can be compared with the $\theta$ angle in QCD, in fact, both provide effects only in non-perurbative quantum regimes. 

Recently Perez and Rovelli on one side \cite{perrov05} and Freidel, Minic and Takeuchi on the other side \cite{fremintak05} showed that when minimally coupled spinor fields are present, the Holst modification is, in general, no more ``topological'' and, as a consequence, the effective theory is not the expected Einstein-Cartan theory. The reason is connected with the well known fact that the presence of fermions, minimally coupled to the gravitational field, modifies the structure of space-time, yielding a non vanishing torsion tensor \cite{sha01}. This implies some modifications in the effective theory involving the Immirzi parameter, thus opening the way to a (classical) physical interpretation of this parameter, which comes out to be related with the coupling constant in front of the four fermions interaction term characterizing the Einstein-Cartan effective theory\footnote{Furthermore, as showed in \cite{fremintak05,ran05}, the introduction of a non minimal coupling in the fermions action leads not only to a modification in the coupling constant in front of the four fermions interaction term, but also to a parity violation in the effective field theory, unfortunately this effect cannot provide constraints on the value of the Immirzi parameter \cite{fremintak05}.}. But this harshly contrasts with the spirit of the Ashtekar-Barbero-Immirzi formalism and with the quantization program of GR: we expect, in fact, that also in the presence of spinor matter, which is a fundamental ingredient of any physical theory, the low energy limit be the ordinary one. In the case of gravity we expect to find the Einstein-Cartan theory as effective low energy limit as in the original Ashtekar-Romano-Tate's paper \cite{ashromtat89} (see also \cite{jac88}), in which the inclusion of bosons and fermions however leads to the Einstein-Cartan as effective theory. In other words we expect that, also in presence of matter, the Immirzi parameter had none effect in the classical theory.  

The request that the classical effective theory be not affected, in any case, by the Immirzi parameter implies a generalization of the Holst covariant approach. Since the presence of spinor fields generates torsion, then we expect that a central role be played by the following term
\begin{equation}\label{nieh-yan}
S_{NY}\left(e,\omega,T\right)=\int\left(T^a\wedge T_a-e_{a}\wedge e_{b}\wedge R^{a b}\right),
\end{equation}
called Nieh-Yan invariant \cite{nieyan82}, which is the only exact 4-form invariant under local Lorentz transformations associated with torsion \cite{chazan97} and represents the natural generalization of the Holst modification to the Hilbert-Palatini action.

In this paper we present a theory where fermions coupled to the gravitational field are present: the gravitational field is described by the Holst action, while fermions are described by a non-minimal Dirac Lagrangian. The non-minimal Lagrangian is introduced so that the low energy effective dynamics be equivalent to the one of the Einstein-Cartan theory, with the typical axial-axial currents interaction. Our theory generalizes the Ashtekar-Romano-Tate one \cite{ashromtat89} (which can be obviously obtained from the general case taking $\beta=\,i$), to arbitrary values of the Immirzi parameter, as a consequence, also to the case of real connections, which plays a central role in the quantization program.

The non-minimal coupling term present in our approach together with the additional term in the Holst action reconstruct the Nieh-Yan invariant (\ref{nieh-yan}), as expected from the very beginning, ensuring that the classical dynamics be the one described by the Einstein-Cartan theory.

The non-minimal spinor action can be, unexpectedly, separated in two independent actions with different weights depending on the Immirzi parameter where the respective interaction terms contain the self-dual and anti-self-dual Ashtekar connections; this suggests to search for a similar separation in the Holst action, in order to rewrite the total action as the sum of two actions describing independently the self-dual and anti-self-dual sector of the complete theory. This separation is in fact possible \cite{ran05-1,ale05} and seems to reflect a partial parity violation in the gravitational interaction\footnote{In the interesting paper \cite{ran05-1} the author shows how to extend the Kodama state to arbitrary values of the Immirzi parameter just using the separation of the Holst action we are speaking about.}.

The plan of the paper is the following one:
\newline In Section \ref{par2} we introduce the Holst action, clarifying the role of the Immirzi parameter.
\newline In Section \ref{par3} we give a brief review of Einstein-Cartan theory, digressing on the main results of \cite{perrov05} and \cite{fremintak05}.
\newline In Section \ref{par4} we motivate and present our approach, moreover the link between the non-minimal coupling term in the spinor action and the Nieh-Yan invariant is explained (a brief description of the Nieh-Yan topological term and its relation with the Pontryagin four dimensional classes are contained in the Appendix \ref{par5}); finally we show that the Ashtekar-Romano-Tate Lagrangian is obtainable by our non-minimal Lagrangian in the limit $\beta=\pm\,i$ (this part of Section \ref{par4} is supported by the formulas and results contained in the Appendix \ref{par6}).
\newline In Section \ref{par8} the separation of the total action in self-dual and anti-self-dual part is carried out.
\newline In Section \ref{par7} we present the canonical Hamiltonian theory, we calculate the constraints of the theory for general values of the Immirzi parameter, starting from the Lagrangian introduced in section \ref{par4}.
\newline A discussion of the result presented is contained in the Concluding Remarks.

Along all the paper we use the sign convention of the Landau Lif$\check{s}$its series. We assume $8\pi G=1$.

\section{The Holst action and the role of the Immirzi parameter}\label{par2}

Let us introduce in this section the Holst action, which represents an important contribute in understanding the geometrical content of the Ashtekar-Barbero-Immirzi formalism. In \cite{hol96}, the author shows that the Barbero's Hamiltonian formulation of General Relativity can be derived from an action which generalizes the ordinary Hilbert-Palatini action. The Holst action is:
\begin{align}\label{holst action}
\nonumber & S_{Ho l}=S_{HP}\left(e,\omega\right)+S_{T T}\left(e,\omega\right)
\\
& =\frac{1}{4}\int\left(\epsilon_{a b c d}\,e^{a}\wedge e^{b}\wedge R^{c d}-\frac{2}{\beta}\,e_{a}\wedge e_{b}\wedge R^{a b}\right),
\end{align}
where $e^{a}$ is the gravitational field, while $R^{a b}=d\omega^{a b}+\omega^{a}_{\phantom1c}\wedge\omega^{c b}$ is the Riemann curvature 2-form and $\omega^{a b}$ is the Lorentz valued spin connection 1-form. The constraints coming out from the ADM 3+1-splitting of the Holst action are the following ones \cite{ashlew04} (Indexes from the beginning of the Greek alphabet $\alpha,\beta,\gamma\dots$ are spatial indexes, while indexes from the middle of the Latin alphabet, $i,j,k\dots$ denote internal degrees of freedom, both the sets of indexes run from $1$ to $3$):
\begin{subequations}\label{constraints}
\begin{align}
&G_i=\mathcal{D}_{\alpha}E^{\alpha}_i=\partial_{\alpha}E^{\alpha}_i+\epsilon_{i j}^{\phantom1\phantom1k}\ ^{(\beta)}\mathcal{A}^{j}_{\alpha}E^{\alpha}_{k}=0 \label{gauss law},
\\
&C_{\alpha}=E_{i}^{\beta}\mathcal{F}^{i}_{\alpha\beta}=0 \label{vector constraint},
\\
&C=\frac{1}{2}E^{\alpha}_{i}E^{\beta}_{j}\,\left[\epsilon^{i j}_{\phantom1\phantom1k}\mathcal{F}^{k}_{\alpha\beta}+2\left(\beta^{2}+1\right)K_{[\alpha}^{i}K_{\beta]}^{j}\right]=0 \label{scalar constraint},
\end{align}
\end{subequations}
where
\begin{subequations}\label{definitions}
\begin{align}
^{(\beta)}\mathcal{A}^i_{\alpha}=&\ \Gamma^i_{\alpha}+\,\beta K^i_{\alpha}
\\
E_i^{\alpha}&=-\frac{1}{2}\,\epsilon_{i j k}\varepsilon^{\alpha\beta\gamma}e^{\phantom1j}_{\beta}e^{\phantom1k}_{\gamma}=-\sqrt{|\det q|}\,e_{\phantom1i}^{\alpha},
\\
& \mathcal{F}^{k}_{\alpha\beta}=2\partial_{[\alpha}\ ^{(\beta)}\mathcal{A}^{k}_{\beta]}+\epsilon^{k}_{\phantom1i j}\ ^{(\beta)}\mathcal{A}_{\alpha}^{i}\ ^{(\beta)}\mathcal{A}_{\beta}^{j},
\end{align}
\end{subequations}
are respectively the connection 1-form taking values in $SU(2)$ or $SO(3)$, its conjugate momentum and the curvature 2-form associated with the connection $A^i$; $\det{q}$ represents the determinant of the metric on the 3-dimensional spatial surface \cite{thi01}. The first class secondary constraints (\ref{constraints}) reflect the gauge freedom of the physical theory, in particular the internal automorphisms of the gauge bundle and the diffeomorphisms invariance of the space-time, constraining the system on a restricted region of the phase space. 
\newline In operating the 3+1-splitting of the Holst action we can appreciate a profound difference between Ashtekar and Barbero connections \cite{sam01}, indeed, while the Ashtekar connections are the pullback to the spatial surface of the 4-dimensional connection, as it appears clear in the Lagrangian formulation presented by Jacobson and Smolin \cite{jacsmo88}; the Barbero ones are defined using specific components of the space-time connection. As a consequence the holonomy along a loop of the Barbero connections depends on the slicing \cite{sam00}, while the holonomy of the Ashtekar connection depends only on the loop. 

As far as the constraints are concerned it is worth noting that the Gauss' law (\ref{gauss law}) and the vector constraint (\ref{vector constraint}) does not depend on $\beta$, while the scalar constraint (\ref{scalar constraint}) is $\beta$-dependent, implying that the physical predictions of the quantum theory will in general depend on the Immirzi parameter; a striking consequence is that even physical quantities not directly depending on the Hamiltonian, for example the area operator, as above remarked, come out to be $\beta$-dependent.

Even though the Immirzi parameter has an important role in the quantum regimes, as just remarked, it plays no role in the classical dynamics, in fact, the Holst action differs from the Hilbert-Palatini action for the presence of the following term:
\begin{equation}\label{top term}
S_{TT}\left(e,\omega\right)=-\frac{1}{2\beta}\int e_{a}\wedge e_{b}\wedge R^{a b},
\end{equation}
which does not affect the classical dynamics; in fact the variation of the Holst action (\ref{holst action}) with respect to the spin connection $\omega^{a b}$ gives us the II Cartan structure equation in the torsion-less case
\begin{equation}\label{cartan equation}
d e^{a}+\omega_{\phantom1b}^{a}\wedge e^{b}=T^a=0,
\end{equation}
this equation implies the following identity 
\begin{equation}\label{bianchi ciclic}
R_{\phantom1b}^{a}\wedge e^{b}=0,
\end{equation}
then the variation of the Holst action with respect to the gravitational field $e^a$ leads to the usual Einstein dynamical equation:
\begin{equation}\label{einstein equations}
\epsilon_{a b c d}\,e^{b}\wedge R^{c d}\left(\omega\left(e\right)\right)=0,
\end{equation}
where we have taken into account the identity in line (\ref{bianchi ciclic}).

It is worth stressing that the presence of a torsion tensor in the right hand side of the II Cartan structure equation would make the Holst action no more dynamically equivalent to the Hilbert-Palatini action. In other words for the action $S_{TT}$ to be a topological term the II Cartan structure equation with vanishing right hand side has to be valid, otherwise the Bianchi identity would assume its general form:
\begin{equation}\label{bianchi ciclic general}
R_{\phantom1b}^{a}\wedge e^{b}=d T^a+\omega^{a}_{\phantom1b}\wedge T^b
\end{equation}
and a contribution proportional to the Immirzi parameter would appear in the right side of the Einstein field equations. 

Then, in vacuum, an analogy between the Immirzi parameter and the $\theta$-angle in QCD exists: their position with respect to the topological term is just the same, moreover both do not affect the classical dynamical vacuum equations and appear only in non-perturbative quantum effects (an interesting work on this subject is \cite{gamobrpul98}, where the authors propose an analogy between the Immirzi ambiguity and ambiguities present in Yang-Mills and Maxwell theories).
It is interesting to ask whether this analogy survives in the presence of spinor matter.

\section{Immirzi parameter and effective theory}\label{par3}

The Einstein-Cartan theory describes a system of fermion fields coupled to gravity, the action for this system is the following one:
\begin{align}\label{einstein-cartan action}
\nonumber S_{EC}\left(e,\omega,\psi,\overline{\psi}\right)=\frac{1}{4} & \int\epsilon_{a b c d}\,e^{a}\wedge e^{b}\wedge R^{c d}
\\
+\frac{i}{2} & \int\star\; e_a\wedge\left(\overline{\psi}\gamma^a\mathcal{D}\psi-\overline{\mathcal{D}\psi}\gamma^a\psi\right),
\end{align}
where the symbol ``$\ \star\ $'' indicates the Hodge dual, while the covariant derivative operator $\mathcal{D}$ acts on the spinor fields as follows:
\begin{equation}\label{covariant derivatives}
\mathcal{D}\psi=d\psi-\frac{i}{4}\;\omega^{a b}\Sigma_{a b}\psi\,\ \ \text{and}\,\ \  \overline{\mathcal{D}\psi}=d\overline{\psi}+\frac{i}{4}\;\overline{\psi}\Sigma_{a b}\omega^{a b},
\end{equation}
$\Sigma_{a b}$ are the generators of the Lorentz group, defined in lines (\ref{gamma matrices}).

In order to get the effective theory, we should vary the action above with respect to the spin connection field $\omega^{a b}$ and solve the II Cartan structure equation resulting from the principle of stationary action, obtaining the expression of the spin connection as function of the gravitational and spinor fields. A useful characteristic of the solution of the II Cartan structure equation is that the pure gravitational contribution to the spin connection can be separated from other possible contribution, this allows us to write
\begin{equation}\label{general spin connection}
\omega^{a b}_{\mu}=\,\stackrel{\circ}{\omega}^{a b}_{\mu}\left(e\right)+K^{a b}_{\mu}, 
\end{equation}
where $\stackrel{\circ}{\omega}^{a}_{\mu b}\left(e\right)=e^{\phantom1a}_{\rho}\nabla_{\mu}e^{\rho}_{\phantom1b}$ is the pure gravitational part of the total spin connection: in other words it is the solution of the II Cartan structure equation in the torsion-less case (\ref{cartan equation}) (the symbol ``$\ \stackrel{\circ}{}\ $'' will denote the torsion-less geometrical objects), while $K^{a b}_{\mu}=K^{\nu\rho}_{\phantom1\phantom2\mu}e_{[\nu}^{\phantom1a} e_{\rho]}^{\phantom1b}$ is the tetrad projection of the so-called contortion tensor $K^{\nu\rho}_{\phantom1\phantom2\mu}$, which is different from zero only when external sources for torsion are present in the total action. Some formulas and definitions are useful for what follows: first of all, we recall the following relation between the contortion tensor and torsion  
\begin{equation}\label{torsion contortion}
K^{\nu}_{\phantom2\rho\mu}=\frac{1}{2}\left(T^{\nu}_{\phantom2\rho\mu}-T^{\phantom1\nu}_{\rho\phantom2\mu}-T^{\phantom1\nu}_{\mu\phantom2\rho}\right),
\end{equation}
it is worth noting that while the contortion tensor is antisymmetric in the first two indexes: $K_{\nu\rho\mu}=-K_{\rho\nu\mu}$, the torsion tensor is antisymmetric in the last two indexes: $T_{\nu\rho\mu}=-T_{\nu\mu\rho}$.
\newline It is useful to divide torsion into its irreducible parts:\\the trace vector
\begin{subequations}\label{TSQ}
\begin{equation}\label{scalar vector}
T_{\mu}=T^{\nu}_{\phantom1\mu\nu};
\end{equation}
the pseudo-trace axial vector
\begin{equation}
S_{\mu}=\epsilon_{\mu\nu\rho\sigma}T^{\nu\rho\sigma}
\end{equation}
and the tensor
\begin{equation}
q_{\nu\rho\sigma},\quad\text{satisfying:}\,\  q^{\nu}_{\phantom1\rho\nu}=0\quad\text{and}\quad\epsilon_{\mu\nu\rho\sigma}q^{\nu\rho\sigma}=0.
\end{equation}
\end{subequations}
The expression of the torsion tensor through the above new fields is:
\begin{equation}
T_{\mu\nu\rho}=\frac{1}{3}\left(T_{\nu}g_{\mu\rho}-T_{\rho}g_{\mu\nu}\right)-\frac{1}{6}\epsilon_{\mu\nu\rho\sigma}S^{\sigma}+q_{\mu\nu\rho},
\end{equation}
where $g_{\mu\nu}=\eta_{a b}e^{\phantom1a}_{\mu}e^{\phantom1b}_{\nu}$ is the metric tensor\footnote{A detailed classification of the torsion components can be found in \cite{caplamsto01}, while multidimensional classical gravity with torsion is described in \cite{helpensha00}.}.
Using the above formulas and definitions we can rewrite the Einstein-Cartan action as follows:
\begin{widetext}
\begin{align}\label{e-c action}
\nonumber S_{EC}\left(e,S,T,q,\psi,\overline{\psi}\right)=-\frac{1}{2}&\int d^4x\det(e)\left(e^{\mu}_{\phantom1a}e^{\nu}_{\phantom1b}\stackrel{\circ}{R}_{\mu\nu}^{\phantom1\phantom2a b}-2\stackrel{\circ}{\nabla}_{\mu}T^{\mu}-\frac{2}{3}\,T_{\rho}T^{\rho}+\frac{1}{24}\,S_{\nu}S^{\nu}+\frac{1}{2}\,q_{\mu\nu\rho}q^{\mu\nu\rho}\right)
\\
+\frac{i}{2}&\int d^4x\det(e)\left(e^{\mu}_{\phantom1a}\left(\overline{\psi}\gamma^a\stackrel{\circ}{\mathcal{D}}_{\mu}\psi-\overline{\stackrel{\circ}{\mathcal{D}}_{\mu}\psi}\gamma^a\psi\right)+\frac{i}{4}\,S_{\rho}J^{\rho}_{(A)}\right),
\end{align}
\end{widetext}
where $J^{\rho}_{(A)}=\overline{\psi}\gamma^{\rho}\gamma^5\psi$ indicates the spinor axial current.
For completeness we wrote in the action also a total divergence term, which can be dropped out supposing the manifold we are integrating over is compact without boundary or requiring that the fields vanish on the boundary.  

In order to calculate the effective theory, we apply the variational principle to the trace vector $T_{\mu}$, pseudo-trace axial vector $S_{\nu}$ and $q_{\mu\nu\rho}$, obtaining respectively the following equations:
\begin{equation}\label{e-c solution}
T^{\mu}=0,\quad \frac{1}{24}S^{\nu}+\frac{1}{8}J^{\nu}_{(A)}=0,\quad q^{\mu\nu\rho}=0;
\end{equation}
the solutions above allows to rewrite the spin connection as
\begin{equation}\label{spin connection 1}
\omega^{a b}_{\mu}=\,\stackrel{\circ}{\omega}\,^{a b}_{\mu}+\frac{1}{4}\epsilon^{a b}_{\phantom1\phantom1c d}e^{c}_{\mu}J^{d}_{(A)},
\end{equation}
which is, as expected, of the form in line (\ref{general spin connection}). Now on the base of a well known theorem\footnote{Let $S\left(q_i,Q_j\right)$ be an action depending on two sets of dynamical variables, $q_i$ and $Q_j$. The solutions of the dynamical equations are extrema of the action with respect to both the two sets of variables: if the dynamical equations $\partial S/\partial q_i=0$ have a unique solution, $q_i^{(0)}\left(Q_j\right)$ for each choice of $Q_j$, then the pullback $S\left(q_i\left(Q_j\right),Q_j\right)$ of the action to the set of solution has the property that its extrema are precisely the extrema of the total total action $S\left(q_i,Q_j\right)$ \cite{ashromtat89}.}, we reinsert the solutions of the equations above into the action obtaining:
\begin{align}\label{j-j action}
\nonumber &S_{J_{(A)}-J_{(A)}}\left(e,\psi,\overline{\psi}\right)=\int d^4x\det(e)\left(-\frac{1}{2}\,e^{\mu}_{\phantom1a}e^{\nu}_{\phantom1b}\stackrel{\circ}{R}_{\mu\nu}^{\phantom1\phantom2a b}\right.
\\
&+\left.\frac{i}{2}\, e^{\mu}_{\phantom1a}\left[\overline{\psi}\gamma^a\stackrel{\circ}{\mathcal{D}}_{\mu}\psi-\overline{\stackrel{\circ}{\mathcal{D}}_{\mu}\psi}\gamma^a\psi\right]+\frac{3}{16}\,\eta_{a b}J_{(A)}^{a}J^{b}_{(A)}\right).
\end{align}
This is the well known effective action of the Einstein-Cartan theory with the peculiar Fermi-like four fermions interacting term. In this approach torsion is a non-dynamical field, in particular the pseudo-trace axial vector is the only interacting field of the irreducible parts of the contortion tensor and is characterized by a contact interaction with the axial spinor current; as a consequence, the effective theory is completely torsion free, as one can easily recognize looking at the action (\ref{j-j action}), but the dynamics is complicated by the presence of a four fermions point-like interaction, ``mediated'' by the non-dynamical pseudo-trace axial vector field. The four fermions term becomes important when the energy available reaches a huge critical density \cite{hehvdh73} and it can be easily neglected at the energy at present available in ``on Earth'' experiments, but it could have important effects in Cosmology, because the very early Universe reached energies even higher than those required to make the $J_{(A)}-J_{(A)}$ term dominant.

Generalizing the Einstein-Cartan theory substituting the Hilbert-Palatini action with the Holst action in line (\ref{holst action}), the dynamical equations for the irreducible components of the torsion tensor change and come out to be dependent on the Immirzi parameter. In particular, being
\begin{align} 
\nonumber\int e^a\wedge e^b\wedge & R_{a b}=\frac{1}{2}\int d^4x\det(e)\left(e^{\mu}_{\phantom1a}e^{\nu}_{\phantom1b}\epsilon^{a b}_{\phantom1\phantom1c d}\stackrel{\circ}{R}^{\phantom1\phantom1c d}_{\mu\nu}\right.
\\
& \left.-\nabla_{\mu}S^{\mu}-\frac{2}{3}T_{\mu}S^{\mu}-\epsilon_{\mu\nu\rho\sigma}q_{\tau}^{\phantom1\mu\rho}q^{\tau\nu\sigma}\right),
\end{align}
the irreducible components of the torsion tensor are given by the following expressions:
\begin{subequations}\label{solutions beta}
\begin{align}\label{solutions beta-1}
&T^{\rho}=\frac{3}{4}\,\frac{\beta}{\beta^2+1}J^{\rho}_{(A)},
\\
S^{\sigma}=-&\frac{3\beta^2}{\beta^2+1}J^{\sigma}_{(A)},\,\ \ \ q^{\mu\nu\rho}=0,
\end{align}
\end{subequations}
and reduce to the ones in line (\ref{e-c solution}) in the limit $\beta\rightarrow\infty$. It is worth noting that the solutions above are not completely consistent, because the solution in line (\ref{solutions beta-1}) gives a relation of proportionality between a vector and a pseudo-vector which have different properties under coordinates transformations, moreover the same relation constraints the Immirzi parameter to be real otherwise the scalar vector irreducible component of torsion would become imaginary (this also prevents by a possible divergence of the action, as one can easily realize looking at the constant in front of the four Fermions interaction term below). However a formal substitution of the above solutions in the action provides the result obtained in \cite{perrov05} and in \cite{fremintak05}:
\begin{align}\label{j-j beta action}
\nonumber S_{J_{(A)}-J_{(A)}}&\left(e,\psi,\overline{\psi}\right)=\int d^4x\det(e)\left(-\frac{1}{2}\,e^{\mu}_{\phantom1a}e^{\nu}_{\phantom1b}\stackrel{\circ}{R}_{\mu\nu}^{\phantom1\phantom2a b}\right.
\\
&+\frac{i}{2}\, e^{\mu}_{\phantom1a}\left(\overline{\psi}\gamma^a\stackrel{\circ}{\mathcal{D}}_{\mu}\psi-\overline{\stackrel{\circ}{\mathcal{D}}_{\mu}\psi}\gamma^a\psi\right)
\\
\nonumber&+\left.\frac{3}{16}\,\frac{\beta^2}{\beta^2+1}\,\eta_{a b}J_{(A)}^{a}J^{b}_{(A)}\right),
\end{align}
where we have taken into account the identity in line (\ref{bianchi ciclic}). This result shows that the Immirzi parameter, which appears in the action for the gravitational field used as starting point in the construction of LQG, not only appears in non-perturbative quantum effects but also in the classical equations of motion, when fermions are present, leading to (independently from the quantum theory) possible observable effects. It plays the role of coupling constant in front of the four fermions interacting term.

In this respect the analogy between the $\theta$-term in QCD and the Holst modification to the Hilbert-Palatini action cannot be put forward in presence of spinor matter, indeed, while the former remains a topological term of the Pontryagin class even though minimally coupled fermion fields are present, the latter is no more a topological term. The presence of spinor matter generates, in fact, a non-vanishing torsion, thus, as formally remarked in the previous section, the Bianchi identity assumes its general form, preventing the Holst action to yield a dynamics completely equivalent to the ordinary one. In particular the resulting action differs from the usual Einstein-Cartan one by a $\beta$ dependent coupling constant in front of the Fermi-like interaction term. We stress that this result, even though interesting, cannot be accepted because of the inconsistency of the relation in line (\ref{solutions beta-1}) and, from a more conceptual point of view, because we expect that the effective dynamics of the theory be completely equivalent to the one coming out from the Einstein-Cartan theory.

In this respect, in the next section we present a theory which has these characteristics: it reduces to the Einstein-Cartan theory once the II Cartan structure equation is solved, it is valid for every values of the Immirzi parameter and in the limit $\beta=\pm\,i$ it provides the Ashtekar-Romano-Tate theory \cite{ashromtat89}.

\section{Effective theory and the Nieh-Yan invariant}\label{par4}

The Immirzi parameter represents an open problem of LQG, which is, of course, one of the most promising approaches to QG. The most striking result about the Immirzi parameter concerns the physical predictions of the quantum theory, which, for a reason explained in \cite{rovthi98}, are affected by the presence of this parameter. Nevertheless the fact that the entropy of a black hole could depend on a classically non-physical parameter is not completely satisfying. An interesting solution to this problem could be found using a gauge group larger than $SU(2)$, as suggested by Immirzi himself in \cite{imm97-1} and developed by Alexandrov, who, in the framework of \emph{Lorentz Covariant Loop Gravity}, shows that no dependence on the Immirzi parameter appears in the spectrum of the area operator \cite{ale00, ale01} (see also \cite{aleliv02} for an overlooking on the drawbacks of the $SU(2)$ formalism). In this way we do not need to give any physical interpretation to the Immirzi parameter, just because it does affect neither the classical nor the quantum sectors of the theory. 

However the origin of this ambiguity is still argument of discussion (an interesting paper on this topic is \cite{chotunyu-05}). In this respect, the attempt to give a classical physical meaning to the Immirzi parameter by minimally coupling spinor fields to gravity has not led to a completely consistent result as showed in the previous section. Thus according to the general idea that inspired the introduction of new variables for General Relativity, we present below a theory containing spinor fields, which, in the framework of the covariant approach to the Ashtekar-Barbero-Immirzi formalism, has as effective limit the usual Einstein-Cartan theory. In other words, we propose an action for the coupled system of spinor and gravitational fields, which contains two modifications with respect the Einstein-Cartan action in line (\ref{einstein-cartan action}): the first is the well known Holst modification, the other one is a non-minimal interaction term in the spinor action. The main result of this section is the reduction of these modifications to the Nieh-Yan invariant (\ref{nieh-yan}), which, as demonstrated in the Appendix \ref{par5}, can be rewritten as a total divergence; moreover, since in the approach we present below the Immirzi parameter results to be a multiplicative factor in front of the Nieh-Yan invariant, then it does not affect the classical dynamical equations. \newline It is worth noting that the Nieh-Yan invariant reduces to a topological invariant on compact manifold (we address the reader to the Appendix \ref{par5} for a brief description of the Nieh-Yan topological term), thus the parallel with the $\theta$-term in QCD can be finally restored.

The action we propose is:
\begin{widetext}
\begin{align}\label{new action}
\nonumber S\left(e,\omega,\psi,\overline{\psi}\right)=\frac{1}{4}\int&\left(\epsilon_{a b c d}\,e^{a}\wedge e^{b}\wedge R^{c d}-\frac{2}{\beta}\,e_{a}\wedge e_{b}\wedge R^{a b}\right)
\\
&+\frac{i}{2}\int\star\; e_a\wedge\left[\overline{\psi}\gamma^a\left(1-\frac{i}{\alpha}\gamma_5\right) \mathcal{D}\psi-\overline{\mathcal{D}\psi}\left(1-\frac{i}{\alpha}\gamma_5\right)\gamma^a\psi\right],
\end{align}
where the covariant derivative operators are defined in line (\ref{covariant derivatives}) and $\alpha$ is a constant. 

Now, using the relations in line (\ref{gamma-sigma}) and the definitions in lines (\ref{TSQ}), we rewrite the action above giving the explicit form of the contributes coming from the irreducible components of the torsion tensor:
\begin{align}\label{new action e-c}
\nonumber &S\left(e,S,T,q,\psi,\overline{\psi}\right)=-\frac{1}{2}\int d^4x\det(e)\,e^{\mu}_{\phantom1a}e^{\nu}_{\phantom1b}\left(\stackrel{\circ}{R}_{\mu\nu}^{\phantom1\phantom2a b}+\frac{1}{2\beta}\,\epsilon^{a b}_{\phantom1\phantom1c d}\stackrel{\circ}{R}^{\phantom1\phantom1c d}_{\mu\nu}\right)
\\
-&\frac{1}{2}\int d^4x\det(e)\left[\stackrel{\circ}{\nabla}_{\mu}\left(-\frac{1}{2\beta}\,S^{\mu}-2T^{\mu}\right)-\frac{2}{3}\,T_{\rho}T^{\rho}-\frac{1}{3\beta}\,T_{\rho}S^{\rho}+\frac{1}{24}\,S_{\nu}S^{\nu}+\frac{1}{2}\,q_{\mu\nu\rho}q^{\mu\nu\rho}-\frac{1}{2\beta}\,\epsilon_{\mu\nu\rho\sigma}q_{\tau}^{\phantom1\mu\rho}q^{\tau\nu\sigma}\right]
\\
\nonumber +&\frac{i}{2}\int d^4x\det(e)e^{\mu}_{\phantom1a}\left(\overline{\psi}\gamma^a\stackrel{\circ}{\mathcal{D}}_{\mu}\psi-\overline{\stackrel{\circ}{\mathcal{D}}_{\mu}\psi}\gamma^a\psi\right)+\frac{1}{2\alpha}\int d^4x\det(e)\stackrel{\circ}{\nabla}_{\mu}\left(\overline{\psi}\gamma_5\gamma^{\mu}\psi\right)+\int d^4x\det(e)\left(\frac{1}{2\alpha}\,T_{\rho}-\frac{1}{8}\,S_{\rho}\right)J^{\rho}_{(A)}.
\end{align}
\\[15pt]
Varying now the action with respect to $T^{\mu},\ S^{\nu},\ q^{\rho\sigma\tau}$, we can calculate the expressions of the irreducible components of torsion, which explicitly are 
\begin{equation}
T^{\rho}=\frac{3}{4\alpha}\left(\frac{\alpha\beta-\beta^2}{\beta^2+1}\right)J^{\rho}_{(A)},\qquad
S^{\sigma}=-\frac{3\beta}{\alpha}\,\frac{\alpha\beta+1}{\beta^2+1}J^{\sigma}_{(A)},\qquad
q^{\mu\nu\rho}=0.
\end{equation}
\end{widetext}
The solutions above obviously depend both on the Immirzi parameter and on the constant $\alpha$, but for $\alpha=\beta$ they reduce to those in line (\ref{e-c solution}), yielding a reduction of the action (\ref{new action e-c}) to the usual effective Einstein-Cartan action (\ref{j-j action}), modulo total derivatives, depending on the Immirzi parameter, which do not affect the classical equations of motion.

This derivation refers to general value of the parameter $\beta=\alpha$: it is worth noting that when the Immirzi parameter gets imaginary values the gravitational and the spinor sectors of the action are non-Hermitian, so that reality conditions are necessary to ensure that the evolution be real. On the other hand, when the Immirzi parameter assumes real values, the action is Hermitian as one can easily verify remembering that $\gamma_5^{\dagger}=\gamma_5$.

The action (\ref{new action}) differs from the usual (\ref{e-c action}) for the presence of additional terms, but the low energy effective dynamics these two different actions provide is exactly the same, because, as we are going to show, the additional terms contained in the action (\ref{new action}) with respect to (\ref{e-c action}) are the elements of the Nieh-Yan invariant (\ref{nieh-yan}). Let us consider the following equality:
\begin{widetext} 
\begin{align}\label{equality}
\nonumber \frac{1}{2\beta}\int&\left[e_{a}\wedge e_{b}\wedge R^{a b}+\star\; e_a\wedge\left(\overline{\psi}\gamma_5\gamma^a\mathcal{D}\psi-\overline{\mathcal{D}\psi}\gamma^a\gamma_5\psi\right)\right]
\\
&=\frac{1}{2\beta}\int\left[e_{a}\wedge e_{b}\wedge\stackrel{\circ}{R}\,^{a b}-\frac{1}{3}\,d V\,T^a\,S_a+q_a\wedge q^a-\star\,e_{a}\wedge K^{a}_{\phantom1d}J_{(A)}^{d}+d\left(\star\,e_{a}J^{a}_{(A)}\right)\right],
\end{align}
\end{widetext}
where we divided the spin connection as $\omega^{a b}=\,\,\stackrel{\circ}{\omega}^{a b}+K^{a b}$, being $K^{a b}$ the contortion 1-form. As soon as one realizes that $\star\,e^b\wedge K^a_{\phantom1b}=T^a d V$, where $T^a$ is the scalar vector (\ref{scalar vector}) and takes into account solutions (\ref{e-c solution}), the equality above reduces to the integral of a total divergence:
\begin{align}\label{equality II}
\nonumber\frac{1}{2\beta}\int&\left[e_{a}\wedge e_{b}\wedge R^{a b}+\star\; e_a\wedge\left(\overline{\psi}\gamma_5\gamma^a\mathcal{D}\psi-\overline{\mathcal{D}\psi}\gamma^a\gamma_5\psi\right)\right]
\\
&\qquad\qquad\qquad\ =\frac{1}{2\beta}\int d\left(T_a\wedge e^{a}\right).
\end{align}
In the last equality we have taken into account the fact that the pseudo-trace axial vector is the only irreducible components of the torsion 2-form different from zero. The term in the right side of the equation above, as showed in the appendix \ref{par5}, is just the total divergence to which the Nieh-Yan topological invariant reduces when the II Cartan structure equation is taken into account, it is worth stressing that the formula in line (\ref{discrete spectrum}) can be applied only in the case we deal with compact (Riemannian) space.

In other words, the additional terms in the action proposed in this paper reconstruct the Nieh-Yan invariant as soon as we take into account the solutions in line (\ref{e-c solution}). On the other hand, having added to the original Einstein-Cartan action terms which reduces to total divergence ensures that the classical dynamics be preserved, just like in the original Holst approach, but with the generalization to the case in which spinor fields are present in the dynamics.

Now we show that the action proposed in line (\ref{new action}) reduces to that in the paper of Ashtekar-Romano-Tate \cite{ashromtat89} when the Immirzi parameter is taken equal to the imaginary unit. For $\beta=i$ indeed we have:
\begin{widetext}
\begin{equation}\label{ART action}
S_{ART}\left(e,\omega,\psi,\overline{\psi}\right)=\frac{1}{4}\int\left(\epsilon_{a b c d}\,e^{a}\wedge e^{b}\wedge R^{c d}+2i\,e_{a}\wedge e_{b}\wedge R^{a b}\right)+\frac{i}{2}\int\star\; e_a\wedge\left[\overline{\psi}\left(1+\gamma_5\right)\gamma^a \mathcal{D}\psi-\overline{\mathcal{D}\psi}\gamma^a\left(1+\gamma_5\right)\psi\right].
\end{equation}
\end{widetext}
Considering the complex self-dual connection\footnote{A tensor field is self-dual if $\frac{1}{2}\,\epsilon^{a b}_{\ \ c d}T^{c d}=i T^{a b}$.}:
\begin{equation}
A_{\mu}^{a b}=\omega_{\mu}^{a b}-\frac{i}{2}\,\epsilon^{a b}_{\phantom1\phantom1c d}\omega_{\mu}^{c d},
\end{equation}
with its associated curvature tensor, $F^{\ \ a b}_{\mu\nu}$, which is related to the curvature of the spin connection $\omega_{\mu}^{a b}$ by the following relation:
\begin{equation} 
F^{\phantom1\phantom1a b}_{\mu\nu}=R^{\phantom1\phantom1a b}_{\mu\nu}-\frac{i}{2}\,\epsilon^{a b}_{\phantom1\phantom1c d}R_{\mu\nu}^{\phantom1\phantom1c d},
\end{equation}
which means that the curvature of the self-dual connection is the self-dual part of the curvature of the spin connection \cite{rov91}, then the action in line (\ref{ART action}) can be rewritten as: 
\begin{align}
\nonumber S_{ART}&\left(e,A,\psi,\overline{\psi}\right)=\frac{i}{2}\int e_{a}\wedge e_{b}\wedge F^{a b}
\\
&+i\int\star\;e_a\wedge\left[\overline{\psi}\gamma^a P_{L}\mathcal{D}\psi-\overline{\mathcal{D}\psi}\gamma^a P_{R}\psi\right],
\end{align}
where we have used the definition of the left and right spinor projectors $P_{L}=\frac{1}{2}\left(1-\gamma^5\right)$ and $P_{R}=\frac{1}{2}\left(1+\gamma^5\right)$. Remembering the explicit form of the Dirac matrices $\gamma^a, \gamma^5$ in chiral representation given in lines (\ref{gamma chiral}) and (\ref{gamma5}), we get for the spinor sector of the action (see Appendix \ref{par6} for further details):
\begin{align}\label{spinoraction}
\nonumber &\ S\left(\sigma,A,\xi,\overline{\xi},\eta,\overline{\eta}\right)=\int d^4x \det(\sigma)\mathcal{L}\qquad\text{with}\qquad
\\
&\mathcal{L}=i\sqrt{2}\,\sigma^{\mu}_{\phantom1A A^{\prime}}\left(\overline{\xi}^{A^{\prime}}\mathcal{D}_{\mu}^{(A)}\xi^{A}-\left(\mathcal{D}_{\mu}^{(A)}\overline{\eta}^{A}\right)\eta^{A^{\prime}}\right),
\end{align}
where $\xi^{A}$ and $\eta^{A^{\prime}}$ are $SL(2,C)$ spinor fields, while $\sigma_{\mu}^{A A^{\prime}}$ are the soldering forms.
The covariant derivative acts only on spinor fields with unprimed indexes and assumes the following form:
\begin{equation}
\mathcal{D}^{(A)}\rho^{A}=d\rho^{A}+\frac{1}{2}\ ^{(+)}A^{k}\sigma^{k\,A}_{\phantom1\phantom1\,B}\rho^{B},
\end{equation}
with
\begin{equation}
\ ^{(+)}A^i=A^{0 i}=\omega^{0 i}-\frac{i}{2}\,\epsilon^{i}_{\phantom1j k}\omega^{j k},
\end{equation}
where $\sigma^{A}_{\ B}$ are the Pauli matrices\footnote{Using relation (\ref{contraction1}) we obtain $\mathcal{D}^{(A)}\rho_{A}=d\rho_{A}-\frac{1}{2}A^{k}\sigma^{k\ C}_{\ A}\rho_{C}$.}. So that the Ashtekar connection results to be the complex gauge field of the local $SU(2)$ rotation in the spinor space. This derivation shows how the Ashtekar formalism allows to encode the full local Lorentz gauge symmetry of the theory in the complex SU(2) valued connections, reducing the phase space of the theory to that of a complex Yang-Mills gauge theory, with additional reality conditions needed to ensure that the dynamics be the one of the real Einstein theory. The fact that the covariant derivative acts only on the unprimed (primed) spinor fields is a consequence of the (anti)self-duality of the connection $A$ and, in general, when the Immirzi parameter is different from $\pm\,i$, we should expect that the action contain covariant derivatives operators acting both on unprimed indexes and on the primed ones.

\section{Left-right separation of gravity-spinors action}\label{par8}

In this section we will show that the spinor action for arbitrary values of the Immirzi parameter contains, as anticipated in the previous section, covariant derivatives of primed and unprimed spinor fields, but unexpectedly these contributions separate in two independent parts. This particular feature of the spinor action suggests that the theory associated to an arbitrary value of the Immirzi parameter can be separated in the sum of two independent self and anti-self dual actions with different weights, depending on the value of the Immirzi parameter itself. The resulting separated action, resembling to a partial parity violating action, can give rise to new investigations about symmetries breaking.

In order to show this let us first of all introduce the following connections 
\begin{subequations}\label{new connections}
\begin{align}\label{connections beta i}
^{(+)}A^{i}_{\mu}:=\, & \omega^{0i}_{\mu}-\frac{i}{2}\epsilon^{i}_{\phantom1j k}\omega_{\mu}^{j k},
\\
^{(-)}A^{i}_{\mu}:=\, & \omega^{0i}_{\mu}+\frac{i}{2}\epsilon^{i}_{\phantom1j k}\omega_{\mu}^{j k}.
\end{align}
\end{subequations}
and their inverse 
\begin{subequations}\label{inverse new connections}
\begin{align}
\omega_{\mu}^{0i} & =\frac{1}{2}\left(^{(+)}A^{i}_{\mu}+\,^{(-)}A^{i}_{\mu}\right),
\\
\omega_{\mu}^{jk} & =\frac{i}{2}\,\epsilon^{j k}_{\phantom1\phantom1i}\left(^{(-)}A^{i}_{\mu}-\,^{(+)}A^{i}_{\mu}\right).
\end{align}
\end{subequations}
These relations allow us to rewrite the spin connection matrix $\omega_{\mu}^{a b}\Sigma_{a b}$ in function of the fields $^{(+)}A^{i}_{\mu}$ and $^{(-)}A^{i}_{\mu}$: using the relation in line (\ref{sim rel}) and the explicit form of the matrices $\alpha^i$ and $\Sigma^i$ given in lines (\ref{alpha}) and (\ref{sigma}), we have:
\begin{align}\label{omega A}
\nonumber\omega_{\mu}^{a b}\Sigma_{a b}= & -2\left(i\omega_{\mu}^{0i}\alpha_i+\frac{1}{2}\epsilon^{i}_{\phantom1j k}\omega_{\mu}^{j k}\Sigma_i\right)
\\
= &\, i \left[\left(\Sigma_i-\alpha_i\right)\,^{(-)}A^{i}_{\mu}+\left(\Sigma_i+\alpha_i\right)\,^{(+)}A^{i}_{\mu}\right].
\end{align}

Some algebra allows us to write the spinor sector of the action (\ref{new action}) as
\begin{align}
\nonumber &S\left(\sigma,A,\xi,\overline{\xi},\eta,\overline{\eta}\right)=\int d^4x \det(\sigma)\mathcal{L}\qquad\text{with}\qquad
\\
&\mathcal{L}=i\sqrt{2}\,\sigma^{\mu}_{\phantom1A A^{\prime}}\left[\theta_{L}\left(\overline{\xi}^{A^{\prime}}\mathcal{D}_{\mu}^{(+)}\xi^{A}-\left(\mathcal{D}_{\mu}^{(+)}\overline{\eta}^{A}\right)\eta^{A^{\prime}}\right)\right.
\\
\nonumber&\qquad\qquad\quad\ \,+\left.\theta_{R}\left(\overline{\eta}^{A}\mathcal{D}_{\mu}^{(-)}\eta^{A^{\prime}}-\left(\mathcal{D}_{\mu}^{(-)}\overline{\xi}^{A^{\prime}}\right)\xi^{A}\right)\right],
\end{align}
where we have introduced the constants $\theta_{L}:=\frac{1}{2}\left(1+\frac{i}{\beta}\right)$ and $\theta_{R}:=\frac{1}{2}\left(1-\frac{i}{\beta}\right)$. The covariant derivatives operators act on primed and unprimed spinor field respectively as
\begin{subequations}
\begin{align}
&\mathcal{D}_{\mu}^{(+)}\rho^{A}=\partial_{\mu}\rho^{A}+\frac{1}{2}\,^{(+)}A^{i}_{\mu}\sigma^{i A}_{\phantom1\phantom1B}\rho^B
\\
&\mathcal{D}_{\mu}^{(-)}\tau^{A^{\prime}}=\partial_{\mu}\tau^{A^{\prime}}+\frac{1}{2}\,^{(-)}A^{i}_{\mu}\sigma^{i A^{\prime}}_{\phantom1\phantom1B^{\prime}}\tau^{B^{\prime}}
\end{align}
\end{subequations}
It is worth noting that the spinor action above reduces to the one in line (\ref{spinoraction}) when the Immirzi parameter is equal to the imaginary unit, $i$, while for $\beta=-i$ it gives the Ashtekar formulation in the case of anti-self dual connections. The action above refers to any value of the Immirzi parameter and results to be the sum with different weights of the ``left'' and ``right'' spinor actions, interacting respectively with the self and anti-self dual connection.

The above ``left-right'' separation of the spinor action suggests to search for a similar behavior in the pure gravitational sector of the action (\ref{new action}). This separation is in fact possible as already demonstrated in \cite{ran05-1,ale05}; here, for the sake of completeness and self-consistency of the paper, we briefly describe the main points of such a procedure, addressing the reader to the cited papers for interesting applications to the quantum theory. 

In order to perform the separation we now define the Dirac valued gravitational field $e_{\mu}$ and spin connections $\omega_{\mu}$ as
\begin{subequations}
\begin{align}
e_{\mu}=e_{\mu}^{\phantom1a}\gamma_a,\quad\text{and}\quad\omega_{\mu}=\omega_{\mu}^{a b}\gamma_{[a}\gamma_{b]}=-i\omega_{\mu}^{a b}\Sigma_{a b},
\end{align}
\end{subequations}
thus, remembering the following formulas:
\begin{subequations}
\begin{align}
&tr\left(\gamma^a\gamma^b\gamma^c\gamma^d\right)=4\left(\eta^{a b}\eta^{c d}-\eta^{a c}\eta^{b d}+\eta^{a d}\eta^{b c}\right),
\\
&i\,tr\left(\gamma^5\gamma^a\gamma^b\gamma^c\gamma^d\right)=4\,\epsilon^{a b c d},
\end{align}
\end{subequations}
direct consequence of the definitions in lines (\ref{gamma chiral}) and (\ref{gamma5definition}), 
the Holst action (\ref{holst action}) can be rewritten as: 
\begin{equation}
\nonumber S\left(e,\omega\right)=\frac{1}{16}\int d^4x\det(e)\,tr\left[\left(1-\frac{i}{\beta}\gamma^5\right)e^{\mu}e^{\nu}R_{\mu\nu}\right].
\end{equation}
Now, since we have
$\frac{1}{2}\left(1-\frac{i}{\beta}\gamma^5\right)=\theta_{L}P_{L}+\theta_{L}P_{R}$,
where $P_{L}$ and $P_{R}$ are respectively the left and right projectors, then it is possible to rewrite the total action in line (\ref{new action}) as follows
\begin{widetext}
\begin{align}\label{L-R action} 
\nonumber S_{L-R}=&-\frac{1}{2}\int d^4x\det(e)\,e^{\mu}_{\phantom1a}e^{\nu}_{\phantom1b}\left(\theta_{L}\,^{(+)}F_{\mu\nu}^{\phantom1\phantom2a b}+\theta_{R}\,^{(-)}F_{\mu\nu}^{\phantom1\phantom2a b}\right)
\\
&+\,i\sqrt{2}\int d^4x\det(e)\,e^{\mu}_{\phantom1a}\sigma^{a}_{\phantom1A A^{\prime}}\left[\theta_{L}\left(\overline{\xi}^{A^{\prime}}\mathcal{D}_{\mu}^{(+)}\xi^{A}-\left(\mathcal{D}_{\mu}^{(+)}\overline{\eta}^{A}\right)\eta^{A^{\prime}}\right)\right.
\\
&\nonumber\qquad\qquad\qquad\qquad\qquad\qquad+\left.\theta_{R}\left(\overline{\eta}^{A}\mathcal{D}_{\mu}^{(-)}\eta^{A^{\prime}}-\left(\mathcal{D}_{\mu}^{(-)}\overline{\xi}^{A^{\prime}}\right)\xi^{A}\right)\right]
\end{align}
\end{widetext}
It is important to note that the left and right action are completely independent, being associated respectively with the dynamics of the self and anti-self dual connections interacting with the left and right components of the spinor fields. 

The next step will be the 3+1 splitting of the above L-R action, which we will carry out in the next section: we believe that an important point will be represented by the demonstration that the Barbero constraints given in lines (\ref{constraints}) are obtainable without any effort from the constraints of the L-R action.

\section{Canonical Approach}\label{par7}

In this section we will carry out a 3+1 decomposition of the action proposed in line (\ref{new action}) in order to construct the Hamiltonian constraints. In this respect, we emphasize that, even though a canonical formulation for the real Barbero connections ($\beta=1$, $A^{i}_{\alpha}=\Gamma^{i}_{\alpha}+K^{i}_{\alpha}$) in presence of fermion fields in second order formalism already exists \cite{thi97-1}\footnote{As far as the quantization of diffeomorphisms invariant theories with fermions is concerned, we address the reader to \cite{bae98,thi97-2}.}, to our knowledge a canonical formulation of the dynamics of the gravitational field coupled to spinor matter for arbitrary values of the Immirzi parameter does not exist in the literature; in this sense we generalize the calculation contained in \cite{ashromtat89,thi97-1} for general values of the Immirzi parameter. Moreover it is important to stress that the resulting constraints are affected by the non-minimal Dirac Lagrangian we start from, which has been motivated before in this paper.

Being aware of the comment by Samuel in \cite{sam01} about the incorrectness of using a ``gauge fixed'' Lagrangian to derive the constraints of the theory, nevertheless taking into account the works of Alexandrov \cite{ale00} and Barros e Sa \cite{bar00}, which demonstrate the correctness of the results obtained by Holst with the gauge fixed Lagrangian \cite{sam01, hol96}, in order to arrive at the canonical formulation, we choose the so called ``time gauge''. This choice simplifies the splitting reducing the tetrad field $e_{\mu}^{\phantom1a}$ and its inverse $e^{\nu}_{\phantom1b}$ to the following form:
\begin{equation}\label{vierbein}
e_{\mu}^{\phantom1a}=\left(
\begin{array}{cc}
N & N^{\alpha}e_{\alpha}^{\phantom1i}
\\
0 & e_{\alpha}^{\phantom1i}
\end{array}
\right)
\qquad\text{and}\qquad
e^{\nu}_{\phantom1b}=\left(
\begin{array}{cc}
\displaystyle\frac{1}{N} & 0
\\
-\displaystyle\frac{N^{\beta}}{N} & e^{\beta}_{\phantom1j}
\end{array}
\right),
\end{equation}
we are indicating with $N$ the Lapse function and with $N^{\alpha}$ the Shift vector; the tetrad basis above reconstructs the ADM decomposed metric, in particular we have:
\begin{align}
\nonumber ds^2&=e_{\mu}^{\phantom1a}e_{a\nu}dx^{\mu}dx^{\nu}
\\
&=N^2dt^2-q_{\alpha\beta}\left(N^{\alpha}dt+dx^{\alpha}\right)\left(N^{\beta}dt+dx^{\beta}\right),
\end{align}
where $q_{\alpha\beta}=e_{\alpha}^{\phantom1i}e_{i\beta}$ is the 3-metric.

Using the expression above for the tetrad fields the Holst action assumes the following (3+1) form:
\begin{align}\label{3+1 action}
\nonumber S_{(3+1)}=-\int d t d^3x\sqrt{q}&\left[F_{\alpha t}^{\phantom1\phantom1 i0}e^{\alpha}_{\phantom1i}\right.
-N^{\beta}
e^{\alpha}_{\phantom1i}F_{\alpha\beta}^{\phantom1\phantom1i0}
\\
&\qquad\ +\left.\frac{1}{2}\,N\,e^{\alpha}_{\phantom1i}e^{\beta}_{\phantom1j}F_{\alpha\beta}^{\phantom1\phantom1ij}
\right],
\end{align}
where we introduced the compact notation 
\begin{equation}\label{compact notation}
F_{\mu\nu}^{\phantom1\phantom2 a b}=R_{\mu\nu}^{\phantom1\phantom1 a b}+\displaystyle\frac{1}{2\beta}\,\epsilon^{a b}_{\phantom1\phantom1c d}R_{\mu\nu}^{\phantom1\phantom1 c d}.
\end{equation} 
The explicit expressions of the tensors contained in the action above can be computed by a very long but straightforward calculation, using the following definitions 
\begin{subequations}
\begin{align}
^{(+)}B^i_{\alpha}:=\omega^{0i}_{\alpha}+\frac{1}{2\beta}\,\epsilon^{i}_{\phantom1j k}\omega^{j k}_{\alpha},
\\
^{(-)}B^i_{\alpha}:=\omega^{0i}_{\alpha}-\frac{1}{2\beta}\,\epsilon^{i}_{\phantom1j k}\omega^{j k}_{\alpha};
\end{align}
\end{subequations}
we obtain:
\begin{subequations}
\begin{align}
\nonumber &F_{\alpha t}^{\phantom1\phantom1i 0}=\partial_t\,^{(+)}B^i_{\alpha}-\partial_{\alpha}\left(\omega_t^{0i}+\frac{1}{2\beta}\,\epsilon^i_{\phantom1j k}\omega^{j k}_t\right)+\omega_{t j}^{i}\,^{(+)}B^j_{\alpha}
\\
&-\frac{\beta^2+1}{2\beta}\,\epsilon^i_{\phantom1j k}\omega_t^{0j}\,^{(-)}B^k_{\alpha}+\frac{\beta^2-1}{2\beta}\,\epsilon^i_{\phantom1j k}\omega_t^{0j}\,^{(+)}B^k_{\alpha},
\end{align}
\begin{align}
\nonumber &F_{\alpha\beta}^{\phantom1\phantom1i 0}=2\partial_{[\beta}\,^{(+)}B^i_{\alpha]}+\frac{1-3\beta^2}{4\beta}\,\epsilon^i_{\phantom1j k}\,^{(+)}B^j_{\alpha}\,^{(+)}B^k_{\beta}
\\
&+\frac{\beta^2+1}{4\beta}\,\epsilon^i_{\phantom1j k}\,^{(-)}B^j_{\alpha}\,^{(-)}B^k_{\beta}+\frac{\beta^2+1}{2\beta}\,\epsilon^i_{\phantom1j k}\,^{(+)}B^j_{[\alpha}\,^{(-)}B^k_{\beta]},
\end{align}
\begin{align}
\nonumber &F_{\alpha\beta}^{\phantom1\phantom1i j}=\frac{\beta^2+1}{\beta}\,\epsilon^{i j}_{\phantom1\phantom1k}\partial_{[\alpha}\,^{(-)}B^k_{\beta]}-\frac{\beta^2-1}{\beta}\,\epsilon^{i j}_{\phantom1\phantom1k}\partial_{[\alpha}\,^{(+)}B^k_{\beta]}
\\[6pt]
\nonumber&+\frac{\beta^2-3}{2}\,^{(+)}B^i_{[\alpha}\,^{(+)}B^j_{\beta]}+\frac{\beta^2+1}{2}\,^{(-)}B^i_{[\alpha}\,^{(-)}B^j_{\beta]}
\\
&-\left(\beta^2+1\right)\,^{(+)}B^{[i}_{[\alpha}\,^{(-)}B^{j]}_{\beta]}.
\end{align}
\end{subequations}
From the above general expression of the $(3+1)$ action we can easily deduce the self and anti-self dual ones, respectively associated to $\beta=\,i$ and $\beta=\,-i$, \footnote{It is worth noting that the following relations hold: $^{(\beta)}\mathcal{A}^{i}_{\alpha}=\beta\,^{(+)}B^{i}_{\alpha}$ and $^{(\beta)}\mathcal{A}^{i}_{\alpha}\stackrel{\beta=i}{\rightarrow}\ ^{(+)}\mathcal{A}^{i}_{\alpha}=\,i\,^{(+)}A^{i}_{\alpha}$, in order not to generate confusion note also the difference between the symbol $A$ and $\mathcal{A}$.} the resulting actions are singular: below we directly write the well known secondary first class constraints for $\beta=i$:
\begin{subequations}\label{constraints beta i}
\begin{align}
&G_i^{(+)}=\mathcal{D}^{(+)}_{\alpha}E^{\alpha}_i=\partial_{\alpha}E^{\alpha}_i+\epsilon_{i j}^{\phantom1\phantom1k}\phantom1^{(+)}\mathcal{A}^{j}_{\alpha}E^{\alpha}_{k}=0\label{gauss law beta i},
\\
&C_{\beta}^{(+)}=E_{i}^{\alpha}\phantom1^{(+)}\mathcal{F}^{i}_{\alpha\beta}=0\label{vector beta i},
\\
&C^{(+)}=\frac{1}{2}\,\epsilon^{i j}_{\phantom1\phantom1k}E^{\alpha}_{i}E^{\beta}_{j}\phantom1^{(+)}\mathcal{F}^{k}_{\alpha\beta}=0 \label{scalar beta i},
\end{align}
\end{subequations}
and for $\beta=-i$:
\begin{subequations}\label{constraints beta -i}
\begin{align}
&G_i^{(-)}=\mathcal{D}^{(-)}_{\alpha}E^{\alpha}_i=\partial_{\alpha}E^{\alpha}_i+\epsilon_{i j}^{\phantom1\phantom1k}\phantom1^{(-)}\mathcal{A}^{j}_{\alpha}E^{\alpha}_{k}=0\label{gauss law beta -i},
\\
&C_{\beta}^{(-)}=E_{i}^{\alpha}\phantom1^{(-)}\mathcal{F}^{i}_{\alpha\beta}=0\label{vector beta -i},
\\
&C^{(-)}=\frac{1}{2}\,\epsilon^{i j}_{\phantom1\phantom1k}E^{\alpha}_{i}E^{\beta}_{j}\phantom1^{(-)}\mathcal{F}^{k}_{\alpha\beta}=0 \label{scalar beta -i},
\end{align}
\end{subequations}
where 
\begin{widetext}
\begin{align}
\nonumber^{(+)}\mathcal{A}^{j}_{\alpha}=i^{(+)}A^{j}_{\alpha}=\Gamma^{j}_{\alpha}+i K^{j}_{\alpha},\quad&\text{and}\quad\phantom1^{(-)}\mathcal{A}^{j}_{\alpha}=-i^{(-)}A^{j}_{\alpha}=\Gamma^{j}_{\alpha}-i K^{j}_{\alpha},
\\
\nonumber^{(+)}\mathcal{F}^{i}_{\alpha\beta}=2\partial_{[\alpha}\phantom1^{(+)}\mathcal{A}^{i}_{\beta]}+\,\epsilon^{i}_{\phantom1j k}\phantom1^{(+)}\mathcal{A}^{j}_{\alpha}\phantom1^{(+)}\mathcal{A}^{k}_{\beta},\quad &\text{and}\quad\phantom1^{(-)}\mathcal{F}^{i}_{\alpha\beta}=2\partial_{[\alpha}\phantom1^{(-)}\mathcal{A}^{i}_{\beta]}+\,\epsilon^{i}_{\phantom1j k}\phantom1^{(-)}\mathcal{A}^{j}_{\alpha}\phantom1^{(-)}\mathcal{A}^{k}_{\beta},
\\
\nonumber E_i^{\alpha}=-\frac{1}{2}\,\epsilon_{i j k}\varepsilon^{\alpha\beta\gamma}&e^{\phantom1j}_{\beta}e^{\phantom1k}_{\gamma}=-\sqrt{|\det q|}\,e_{\phantom1i}^{\alpha},
\end{align}
\end{widetext}
are respectively the self and anti-self dual Ashtekar connections, their associated curvature tensors and the conjugate momentum.

Now, in order to write the constraints for the total action (\ref{new action}) we use the result obtained in the previous section. In particular denoting with $\overline{\pi},\ \omega,\ \overline{\omega},\ \pi$ the conjugate momenta associated respectively to the spinor fields $\xi,\ \overline{\eta},\ \eta,\ \overline{\xi}$ the constraints for the L-R action in line (\ref{L-R action}) are
\begin{subequations}\label{constraints general}
\begin{align}
G_i^{(+)}&=-\frac{1}{2}\left(\overline{\pi}\sigma_i\xi+\overline{\eta}\sigma_i\omega\right), \label{gauss1}
\\
G_i^{(-)}&=-\frac{1}{2}\left(\overline{\omega}\sigma_i\eta+\overline{\xi}\sigma_i\pi\right), \label{gauss2}
\end{align}
\begin{align}
\nonumber\theta_{L}C_{\beta}^{(+)}&-\theta_{R}C_{\beta}^{(-)}
\\
&=i\theta_{L}\left(\overline{\pi}\mathcal{D}^{(+)}_{\beta}\xi+\mathcal{D}^{(+)}_{\beta}\overline{\eta}\omega\right)
\\
\nonumber&+i\theta_{R}\left(\overline{\omega}\mathcal{D}^{(-)}_{\beta}\eta+\mathcal{D}^{(-)}_{\beta}\overline{\xi}\pi\right),
\end{align}
\begin{align}
\nonumber\theta_{L}C^{(+)}&-\theta_{R}C^{(-)}
\\
&=\theta_{L}E^{\alpha}_i\left(\overline{\pi}\sigma^i\mathcal{D}^{(+)}_{\alpha}\xi+\mathcal{D}^{(+)}_{\alpha}\overline{\eta}\sigma^i\omega\right)
\\
\nonumber&+\theta_{R}E^{\alpha}_i\left(\overline{\omega}\sigma^i\mathcal{D}^{(-)}_{\alpha}\eta+\mathcal{D}^{(-)}_{\alpha}\overline{\xi}\sigma^i\pi\right).
\end{align}
\end{subequations}
Let us define the new connection
\begin{equation}\label{barbero connection}
^{(\beta)}\mathcal{A}^i_{\alpha}=\frac{1}{\theta_{L}-\theta_{R}}\left(\theta_{L}\phantom1^{(+)}\mathcal{A}^i_{\alpha}-\theta_{R}\phantom1^{(-)}\mathcal{A}^i_{\alpha}\right)
\end{equation}
and the new curvature tensor
\begin{align}\label{barbero curvature tensor}
\nonumber^{(\beta)}\mathcal{F}^i_{\alpha\beta}&=\frac{1}{\theta_{L}-\theta_{R}}\left(\theta_{L}\phantom1^{(+)}\mathcal{F}^i_{\alpha\beta}-\theta_{R}\phantom1^{(-)}\mathcal{F}^i_{\alpha\beta}\right)
\\
&+\left(\beta^2+1\right)\epsilon^{i}_{\phantom1j k}\omega^{0j}_{[\alpha}\omega^{0k}_{\beta]},
\end{align}
we recognize in the connection defined in line (\ref{barbero connection}) the Ashtekar-Barbero-Immirzi connection $^{(\beta)}\mathcal{A}^i_{\alpha}=\Gamma^i_{\alpha}+\beta K^i_{\alpha}$, which result to be a weighted sum of the self and anti-self dual Ashtekar connections.

Subtracting the constraint (\ref{gauss2}) from the (\ref{gauss1}), we obtain the generalization of the Gauss law in the presence of spinor fields and for arbitrary values of the Immirzi parameter: 
\begin{align}\label{gauss general}
&\mathcal{D}_{\alpha}E^{\alpha}_i=\partial_{\alpha}E^{\alpha}_i+\epsilon_{i j}^{\phantom1\phantom1k}\phantom1^{(\beta)}\mathcal{A}^{j}_{\alpha}E^{\alpha}_{k}
\\
\nonumber&=\frac{1}{2}\frac{\theta_{L}}{\theta_{R}-\theta_{L}}\left(\overline{\pi}\sigma_i\xi+\overline{\eta}\sigma_i\omega\right)+\frac{1}{2}\frac{\theta_{R}}{\theta_{L}-\theta_{R}}\left(\overline{\omega}\sigma_i\eta+\overline{\xi}\sigma_i\pi\right).
\end{align}
Reabsorbing the two components spinor fields in the four components Dirac spinor as $\psi=\left(\begin{array}{c}\xi\\\eta\end{array}\right)$ and $\overline{\psi}=\psi^{+}\gamma_0=\left(\overline{\eta}\ \  \overline{\xi}\right)$ and defining the conjugate momenta respectively to $\psi$ and $\overline{\psi}$ as $\Pi=i\sqrt{q}\,\psi^{+}=\left(\overline{\pi}\ \  \overline{\omega}\right)$ and $\Omega=-i\sqrt{q}\,\gamma^0\psi=\left(\begin{array}{c}\omega\\\pi\end{array}\right)$, the Gauss law above can be rewritten in the more readable and compact form:
\begin{align}
\nonumber\mathcal{D}_{\alpha}E^{\alpha}_i&=\partial_{\alpha}E^{\alpha}_i+\epsilon_{i j}^{\phantom1\phantom1k}\phantom1^{(\beta)}\mathcal{A}^{j}_{\alpha}E^{\alpha}_{k}
\\
&=-\frac{i}{4}\left[\Pi\left(\beta\alpha_i+i\Sigma_i\right)\psi+\overline{\psi}\left(\beta\alpha_i+i\Sigma_i\right)\Omega\right].
\end{align}
In order to rewrite the total 3-diffeomorphisms and scalar constraints we have to take into account either the relation in line (\ref{barbero curvature tensor}) either the following ones:
\begin{subequations}\label{scalar derivative}
\begin{align}
\theta_{L}D^{(+)}_{\alpha}=\theta_{L}D^{(\beta)}_{\alpha}-\frac{i}{4}\frac{\beta^2+1}{\beta^2}\Gamma^{i}_{\alpha}\sigma^i,
\\
\theta_{R}D^{(-)}_{\alpha}=\theta_{R}D^{(\beta)}_{\alpha}+\frac{i}{4}\frac{\beta^2+1}{\beta^2}\Gamma^{i}_{\alpha}\sigma^i,
\end{align}
\end{subequations}
we obtain
\begin{widetext}
\begin{align}\label{diffeomorphisms general}
\nonumber E_{i}^{\alpha}\phantom1^{(\beta)}&\mathcal{F}^{i}_{\alpha\gamma}-\frac{\beta^2+1}{\beta^2}\epsilon^{i}_{\phantom1j k}E_{i}^{\alpha}\left(^{(\beta)}\mathcal{A}^j_{\alpha}-\Gamma^j_{\alpha}\right)\left(^{(\beta)}\mathcal{A}^k_{\gamma}-\Gamma^k_{\gamma}\right)
\\
=&\theta_{L}\left(\overline{\pi}\mathcal{D}^{(\beta)}_{\gamma}\xi+\mathcal{D}^{(\beta)}_{\gamma}\overline{\eta}\omega\right)+\theta_{R}\left(\overline{\omega}\mathcal{D}^{(\beta)}_{\gamma}\eta+\mathcal{D}^{(\beta)}_{\gamma}\overline{\xi}\pi\right)-\frac{i}{4}\frac{\beta^2+1}{\beta^2}\Gamma^{i}_{\gamma}\left(\overline{\pi}\sigma^i\xi+\overline{\eta}\sigma^i\omega-\overline{\omega}\mathcal\sigma^i\eta+\overline{\xi}\sigma^i\pi\right).
\end{align}
We note that the second term on the left hand side of the equations above does not vanish trivially as in vacuum, in fact when spinor matter is present the spin connection contains also contributes proportional to spinor field, its general expression is given in line (\ref{spin connection 1}); on the other side it vanishes together with the last term in the same equation when the Immirzi parameter is equal to $\pm\,i$, reducing, respectively, to the 3-diffeomorphisms constraints of the self and anti-self dual theory in presence of spinor fields.

It remains to calculate the scalar constraint. Using as above the relations in lines (\ref{barbero curvature tensor}) and (\ref{scalar derivative}), we obtain
\begin{align}\label{scalar general}
\nonumber \frac{i}{2\beta}E^{\alpha}_{i}E^{\beta}_{j}&\left[\epsilon^{i j}_{\phantom1\phantom1k}\phantom1^{(\beta)}\mathcal{F}^{k}_{\alpha\beta}+2\,\frac{\beta^2+1}{\beta^2}\left(^{(\beta)}\mathcal{A}^i_{\alpha}-\Gamma^i_{\alpha}\right)\left(^{(\beta)}\mathcal{A}^j_{\beta}-\Gamma^j_{\beta}\right)\right]
\\
=\theta_{L}E^{\alpha}_i&\left(\overline{\pi}\sigma^i\mathcal{D}^{(\beta)}_{\alpha}\xi+\mathcal{D}^{(\beta)}_{\alpha}\overline{\eta}\sigma^i\omega\right)+\theta_{R}E^{\alpha}_i\left(\overline{\omega}\sigma^i\mathcal{D}^{(\beta)}_{\alpha}\eta+\mathcal{D}^{(\beta)}_{\alpha}\overline{\xi}\sigma^i\pi\right)
\\
\nonumber & -\frac{i}{4}\frac{\beta^2+1}{\beta^2}E^{\alpha}_i\Gamma^{j}_{\alpha}\left(\overline{\pi}\sigma^i\sigma_j\xi+\overline{\eta}\sigma^i\sigma_j\omega-\overline{\omega}\mathcal\sigma^i\sigma_j\eta+\overline{\xi}\sigma^i\sigma_j\pi\right).
\end{align}
\end{widetext}
The constraints in lines (\ref{gauss general}), (\ref{diffeomorphisms general}) and (\ref{scalar general}) have complicate expressions, on the other side the ones in lines (\ref{constraints general}), besides being polynomial have of course simpler form and could represent an useful tool in view of quantum theory, but since the dynamics is described by the self and anti-self dual Ashtekar variables, additional reality conditions are needed\footnote{As far as the problem of choosing appropriate reality conditions for the left-right separated theory is concerned, we address the reader to \cite{ale05}.}.

\section{Concluding Remarks}\label{conclusion}

In Ashtekar-Barbero-Immirzi formulation of General Relativity the $\beta$-parameter does not play any role in the classical dynamics, in fact it is introduced in the theory via a canonical transformation of the conjugated canonical variables. But since the canonical transformation cannot be implemented unitarily in the quantum theory, then the Immirzi parameter has important effects in the quantum theory leading, as showed by Immirzi himself, to a dependence of the spectra of the quantum operators on the $\beta$ parameter \cite{imm97-1}. As clarified by the Holst covariant formulation, the Immirzi parameter is a multiplicative factor in front of a ``on shell'' topological term, where the on shell condition is represented by the II Cartan structure equation, with vanishing right hand side, i.e. vanishing torsion; in this paper we have shown that the Immirzi parameter plays the same role also in the presence of spinor matter fields, on condition that we consider a non-minimal action for fermion fields. The introduction of the non-minimal action has been motivated by the request that the effective theory, coming out from the coupled gravity-spinor system, be the one of the Einstein-Cartan theory: the demonstration that the sum of the additional Holst modification to the Hilbert-Palatini action and of the non minimal term in Dirac Lagrangian reduces to the Nieh-Yan  invariant provides a geometrical foundation to the theory.

The action we have proposed reduces to the Ashtekar-Romano-Tate action when the Immirzi parameter is equal to the imaginary unit, explaining why the Ashtekar variables were originally introduced in spinor formalism: because they are the natural connection for the left part of a Dirac spinor field.

Our approach generalizes the Ashtekar-Romano-Tate one to the case of real connection; we think this development of the formalism could be important in view of the quantum theory, in fact the difficulties found in implementing the reality conditions at the quantum level have led in these last years to give up the relative simple constraints of the original Ashtekar approach, in favor of real connections, which do not require any additional condition.

A striking feature of the non-minimal spinor action is that it can be rewritten as a sum of two term with different weights depending on the Immirzi parameter and respectively coupled to the self and anti-self dual Ashtekar connections.
We have demonstrated that a similar separation works for the gravitational sector too, this fact simplifies very much the calculation of the constraints of the complete theory, which can be rewritten as a weighted sum of the well known constraints of the self and anti-self dual Ashtekar theory. 

The separation of the action in, let's say, left and right part not only represents a simplification of the constraints, but, since the separated action resembles to the one of a partially parity violating system, it could be an important tool to get an insight into fundamental symmetries of space-time.

Finally, since it exists a dualism between physical (material) frame and time in quantum gravity \cite{mermon03}, the consistent introduction of matter fields in QG could represents a way to solve the problem of time, as many authors have argued during the years \cite{kuctor91, bickuc97, rov91-1, mermon04}.

\appendix\section{Nieh-Yan topological term}\label{par5}

The Nieh-Yan topological term is the only Lorentz invariant exact 4-form including torsion. In fact, using the expression in line (\ref{bianchi ciclic general}) we can write
\begin{align} 
\nonumber e_a\wedge e_b&\wedge R^{a b}-T^a\wedge T_a
\\
=\,&e_a\wedge d T^a+\omega^b_a\wedge e^a\wedge T_b-T^a\wedge T_a,
\end{align}
now by the II Cartan structure equation (\ref{cartan equation}) we have
\begin{align} 
\nonumber e_a\wedge d T^a+\omega^b_a\wedge e^a\wedge T_b&-T^a\wedge T_a
\\
&=e_a\wedge d T^a-d e^b\wedge T_b,
\end{align}
finally we arrive at the expected result
\begin{equation}\label{show}
e_a\wedge e_b\wedge R^{a b}-T^a\wedge T_a=d\left(e_a\wedge T^a\right).
\end{equation}

The integral of the Nieh-Yan topological term over a compact manifold has a discrete spectrum, just like the Euler and Pontryagin classes. It is possible to show in particular that
\begin{align}\label{discrete spectrum} 
\nonumber \int_{M^4}&\left(e_a\wedge e_b\wedge R^{a b}-T^a\wedge T_a\right)
\\
&=\frac{L^2}{2}\left[P_4\left(SO(4)\right)-P_4\left(SO(5)\right)\right],
\end{align}
where with $P_4$ we are indicating the four dimensional Pontryagin classes: 
\begin{equation} 
P_4=\int_{M^4}R^{a b}\wedge R_{a b}.
\end{equation}
To show what stated above it is necessary to embed the group of rotations on the tangent space $SO(4)$ into $SO(5)$ \cite{chazan97}. In order to do this, let us construct a connection for $SO(5)$ combining the spin connection and the tetrad fields:
\begin{equation}
\Omega^{A B}=
\left(
\begin{array}{cc}
	\omega^{a b} & \displaystyle\frac{1}{L}\,e^a 
	\\[12pt]
	-\displaystyle\frac{1}{L}\,e^b & 0
\end{array}
\right)
\end{equation}
where capital letters run from $0$ to $4$, while the constant $L$ has the dimension of a length and is introduced for dimensional reasons. Once constructed the curvature 2-form $F^{A B}$ associated with the connection $\Omega^{A B}$ it is simple to realize what follows:
\begin{equation}
F^{A B}\wedge F_{A B}=R^{a b}\wedge R_{a b}+\frac{2}{L^2}\left(T^a\wedge T_a-e_a\wedge e_b\wedge R^{a b}\right).
\end{equation}
So we have:
\begin{align}
\nonumber \int_{M^4}&\left(e_a\wedge e_b\wedge R^{a b}-T^a\wedge T_a\right)
\\
&=\frac{L^2}{2}\int_{M^4}\left(R^{a b}\wedge R_{a b}-F^{A B}\wedge F_{A B}\right),
\end{align}
which gives the formula (\ref{discrete spectrum}).

\section{Spinor formalism and Dirac equation}\label{par6}

In this appendix the reader can find a brief description of spinor formalism  \cite{wal84, lanlif99}, which will allow us to introduce the spinor representation of Dirac's matrices: many of the expressions and formulas contained in this appendix are useful for the derivation described in the last part of Section \ref{par4} and in Section \ref{par8}.

In relativistic theory two kinds of spinor fields exist. They are indicated with primed $A^{\prime},B^{\prime},C^{\prime}$ and unprimed spinor indexes $A,B,C$, to raise and lower unprimed spinor indexes we use the non-degenerate antisymmetric tensor $\epsilon^{A B}$ and $\epsilon_{A B}$, which following the standard convention \cite{wal84} is defined as minus the inverse of $\epsilon_{A B}$, in other words we have:
\begin{equation}\label{minus sign}
\epsilon^{A B}\epsilon_{B C}=-\delta^{A}_{\phantom1C}.
\end{equation}
Let $W$ be a two dimensional vector space over the complex number, the pair $\left(W,\epsilon_{A B}\right)$ is called a spinor space: the antisymmetric tensor plays the same role of the metric field on the space-time. In order to compensate the minus sign in line (\ref{minus sign}), to raise a spinor index we contract the second index of the antisymmetric tensor, i.e. 
\begin{equation}\label{contraction}
\lambda^{A}=\epsilon^{A B}\lambda_{B}=-\epsilon^{B A}\lambda_{B},
\end{equation}
it is worth paying attention to what follows:
\begin{equation}\label{contraction1}
\mu_{A}\lambda^{A}=\epsilon^{B A}\mu^{B}\lambda^{A}=-\epsilon^{A B}\mu^{B}\lambda^{A}=-\mu^{B}\lambda_{B},
\end{equation}
in particular, for any spinor field $\lambda^{A}$, we have $\lambda^{A}\lambda_{A}=0$.

The request of relativistic invariance fixes for spinor fields the following system of equations:
\begin{subequations}
\begin{align}\label{spinorial equations}
&i\sigma_{\phantom1B^{\prime}A}^a\partial_a\xi^{A}=\frac{m}{\sqrt{2}}\,\eta_{B^{\prime}},
\\
&i\sigma^{\phantom1AB^{\prime}}_a\partial^a\eta_{B^{\prime}}=\frac{m}{\sqrt{2}}\,\xi^{A},
\end{align}
\end{subequations}
where $\sigma^{\phantom1AB^{\prime}}_a$ are the soldering forms: these represent the Dirac equation in spinor representation.
Let us also write down the equations satisfied by the complex conjugate spinor fields $\overline{\xi}$ and $\overline{\eta}$:
\begin{subequations}\label{spinorial complex equations}
\begin{align}
&i\partial_a\overline{\xi}^{A^{\prime}}\sigma_{\phantom1A^{\prime}B}^a=-\frac{m}{\sqrt{2}}\,\overline{\eta}_{B},
\\
&i\partial^a\overline{\eta}_{B}\sigma^{\phantom1BA^{\prime}}_a=-\frac{m}{\sqrt{2}}\,\overline{\xi}^{A^{\prime}},
\end{align}
\end{subequations}
where we used the fact that the soldering forms are Hermitian.
Settling the spinor fields $\xi$ and $\eta$ in the bi-spinor $\psi$ as $\psi=\left(\begin{array}{c}\xi\\\eta\end{array}\right)$, the Dirac equations can be rewritten in the ordinary symmetric form
\begin{equation}\label{Dirac equation}
i\gamma^{\mu}\partial_{\mu}\psi=m\psi,
\end{equation}
the Dirac matrices in spinor (or chiral) representation assume the following form
\begin{equation}\label{gamma chiral}
\gamma^{0}=\left(
\begin{array}{cc}
0 & 1
\\
1 & 0
\end{array}
\right),
\qquad
\gamma^{i}=\left(
\begin{array}{cc}
0 & -\sigma^i
\\
\sigma^i & 0
\end{array}
\right),
\end{equation}
where $\sigma^i$ are the Pauli matrices. In the same representation, introducing the bi-spinor field $\overline{\psi}$ defined as $\overline{\psi}=\psi^{+}\gamma_0=\left(\overline{\eta}\ \ \ \overline{\xi}\right)$, equations (\ref{spinorial complex equations}) can be rewritten in the ordinary symmetric form as
\begin{equation}\label{complex Dirac equation}
i\partial_{\mu}\overline{\psi}\gamma^{\mu}=-m\overline{\psi}.
\end{equation}

It is worth noting that in the low energy limit Dirac equation must reduce to the dynamical equation of spin-$\frac{1}{2}$ non-relativistic particles, in other words we expect that the four components spinor description provided by the Dirac equation reduces to the two components of the non-relativistic Pauli's theory. The limit $p_0\rightarrow m$ and $\mathbf{p}\rightarrow 0$ in equations (\ref{spinorial equations}), in fact, yields the relation $\xi=\eta$, which means that only two of the four components contained in the bi-spinor $\psi$ are independent, so the dynamics is effectively described by a two components spinor field. 

By the way exploiting a particular form of the Dirac equation where an explicit time derivative term is present
\begin{equation}
i\partial_{0}\psi=\mathcal{H}\psi,
\end{equation}
with the Hamiltonian $\mathcal{H}$ defined as $\mathcal{H}=i\alpha^k\partial_k+\beta m$, we have the occasion to introduce the $4\times 4$ matrices $\alpha^i$ and $\beta$, which, in chiral representation, assume the explicit form:
\begin{equation}\label{alpha}
\alpha^i=\gamma^{0}\gamma^i=\left(
\begin{array}{cc}
\sigma^i & 0
\\
0 & -\sigma^i
\end{array}
\right),
\qquad
\beta=\gamma^{0}.
\end{equation}
The matrices $\alpha^i$ are the generators of the Lorentz boosts, indeed in chiral representation for an infinitesimal boost transformation corresponding to the variation of velocity of the reference system $\delta V^i$, the primed and unprimed spinor fields transform as follows:
\begin{equation}
\eta^{\prime}=\left(1+\frac{1}{2}\,\sigma^i\delta V_i\right)\eta\quad\text{and}\quad\xi^{\prime}=\left(1-\frac{1}{2}\,\sigma^i\delta V_i\right)\xi,
\end{equation}
which, using the bi-spinor field, can be rewritten in a more compact form as
\begin{equation}\label{boost}
\psi^{\prime}=\left(1-\frac{1}{2}\,\alpha^i\delta V_i\right)\psi.
\end{equation}
Likewise  for an infinitesimal rotation we have
\begin{equation}\label{rotation}
\psi^{\prime}=\left(1+\frac{i}{2}\,\Sigma^i\delta\theta_i\right)\psi,
\end{equation}
where in the formal vector $\delta\theta^i$ we contain the infinitesimal angles of rotation with respect the three reference axes\footnote{The formula (\ref{boost}) and (\ref{rotation}) are valid in any representation, whether $\alpha^i$ and $\Sigma^i$ indicate the matrices in that specific representation.}.

It is simple to realize that either in the chiral either in the Dirac representation the matrix $\frac{1}{2}\Sigma^{i}$, where
\begin{equation}\label{sigma}  
\Sigma^i=\left(
\begin{array}{cc}
\sigma^i & 0
\\
0 & \sigma^i
\end{array}
\right),
\end{equation}
represents the 3-dimensional spin operator. We remember that in a generic representation this operator can be written as
$-\alpha^i\gamma^5$, this allows us to introduce the explicit form of the Hermitian matrix $\gamma_5$ \cite{lanlif99}
\begin{align}\label{gamma5definition}
\nonumber \gamma^5=&-i\gamma^0\gamma^1\gamma^2\gamma^3=\frac{i}{4!}\epsilon_{a b c d}\gamma^a\gamma^b\gamma^c\gamma^d
\\
=&\,\gamma_5=i\gamma_0\gamma_1\gamma_2\gamma_3=i\gamma^{3\dagger}\gamma^{2\dagger}\gamma^{1\dagger}\gamma^{0\dagger}=\gamma^{5\dagger},
\end{align}
which in the chiral and Dirac representation respectively writes:
\begin{equation}\label{gamma5}
\gamma^5=\left(
\begin{array}{cc}
-1 & 0
\\
0 & 1
\end{array}
\right)
\quad\text{and}\quad
\gamma^5=\left(
\begin{array}{cc}
0 & -1
\\
-1 & 0
\end{array}
\right).
\end{equation}

A generic Lorentz transformation can be written in a compact form introducing the antisymmetric matrix tensor $\Sigma^{a b}$ defined as 
\begin{subequations}\label{gamma matrices}
\begin{align}
& \Sigma^{a b}=\frac{i}{2}\left[\gamma^a,\gamma^b\right],
\\
& i\Sigma^{a b}=\eta^{a b}-\gamma^a\gamma^b;
\end{align}
\end{subequations}
which satisfies the following useful relation:
\begin{align}\label{gamma-sigma}
\left\{\gamma^a,\Sigma^{b c}\right\}&=2\epsilon^{a b c}_{\phantom1\phantom1\phantom1d}\gamma_5\gamma^d,
\\
\left[\gamma^a,\Sigma^{b c}\right]&=4\,i\eta^{a[b}\gamma^{c]}.
\end{align}
The explicit form of the antisymmetric tensor $\Sigma^{a b}$ is:
\begin{equation}\label{sigma explicit}
\Sigma^{a b}=\left(
\begin{array}{cccc}
0 & i\alpha^1 & i\alpha^2 & i\alpha^3
\\
-i\alpha^1 & 0 & \Sigma^3 & -\Sigma^2
\\
-i\alpha^2 & -\Sigma^3 & 0 & \Sigma^1
\\
-i\alpha^3 & \Sigma^2 & -\Sigma^1 & 0
\end{array}\right)=\left(i\alpha^i,-\Sigma^j\right);
\end{equation}
introducing also the infinitesimal antisymmetric tensor $\delta\epsilon^{a b}=\left(\delta V^i,\delta\theta^j\right)$, where for the listing of the components we used the same convention in (\ref{sigma explicit}), then we obtain
\begin{equation}\label{sim rel}
\Sigma^{a b}\delta\epsilon_{a b}=-2\Sigma^i\delta\theta_i-2i\alpha^i\delta V_i,
\end{equation}
consequently we can summarize the transformation laws of spinor fields in lines (\ref{boost}) and (\ref{rotation}) in the following compact form:
\begin{equation}
\psi^{\prime}=\left(1-\frac{i}{4}\Sigma^{a b}\delta\epsilon_{a b}\right)\psi,
\end{equation}
which is the transformation law of a bi-spinor field under a 4-dimensional Lorentz rotation.

\end{document}